# Metabolite essentiality elucidates robustness of *Escherichia coli* metabolism

**Pan-Jun Kim[1,2], Dong-Yup Lee[3−5,a], Tae Yong Kim[1,4,5], Kwang Ho Lee[1,4,5,b], Hawoong Jeong[1−3], Sang Yup Lee[1,3−6], and Sunwon Park[3,5]**

[1]Center for Systems and Synthetic Biotechnology, Institute for the BioCentury, [2]Department of Physics, [3]Bioinformatics Research Center, [4]Metabolic and Biomolecular Engineering National Research Laboratory, BioProcess Engineering Research Center, [5]Department of Chemical and Biomolecular Engineering (BK21 Program), [6]Department of BioSystems, Korea Advanced Institute of Science and Technology, 373-1 Guseong-dong, Yuseong-gu, Daejeon 305-701, Korea | Present address: [a]Department of Chemical and Biomolecular Engineering, National University of Singapore, 4 Engineering Drive 4, Singapore 117576; Bioprocessing Technology Institute, Agency for Science, Technology and Research (A*STAR), 20 Biopolis Way, #06-01, Centros, Singapore; [b]R & D Center for Bioproducts, CJ Corporation, Seoul 157-724, Korea

Correspondence: H.J. (hjeong@kaist.ac.kr) or S.Y.L. (leesy@kaist.ac.kr).

**Complex biological systems are very robust to genetic and environmental changes at all levels of organization. Many biological functions of *Escherichia coli* metabolism can be sustained against single-gene or even multiple-gene mutations by using redundant or alternative pathways. Thus, only a limited number of genes have been identified to be lethal to the cell. In this regard, the reaction-centric gene deletion study has a limitation in understanding the metabolic robustness. Here, we report the use of flux-sum, which is the summation of all incoming or outgoing fluxes around a particular metabolite under pseudo-steady state conditions, as a good conserved property for elucidating such robustness of *E. coli* from the metabolite point of view. The functional behavior, as well as the structural and evolutionary properties of metabolites essential to the cell survival, was investigated by means of a constraints-based flux analysis under perturbed conditions. The essential metabolites are capable of maintaining a steady flux-sum even against severe perturbation by actively redistributing the relevant fluxes. Disrupting the flux-sum maintenance was found to suppress cell growth. This approach of analyzing metabolite essentiality provides insight into cellular robustness and concomitant fragility, which can be used for several**

**applications, including the development of new drugs for treating pathogens.**

Availability of the complete genome sequences for well-characterized organisms has led to the reconstruction of genome-scale metabolic networks, which represent a complex web of metabolites and their interconversions catalyzed by the gene products. Robustness, the inherent property of metabolic networks, enables the maintenance of cellular functions under various internally and externally perturbed conditions. This robustness has been experimentally demonstrated such that even the disruption of a considerable portion of genes could not affect the cell viability (1, 2). Although studies on the topological and functional properties of metabolic networks have achieved much progress (3–6), they still provide only a limited understanding of metabolic robustness. The conventional attempt to study such robustness relies on the identification of the genes or reactions indispensable to a cell. However, universal metabolic pathways across species, such as the tricarboxylic acid (TCA) cycle or glycolytic pathways, have relatively few lethal reactions (1, 7). This fact indicates that the more important a reaction is, the higher the chance is to have a backup pathway (7). Thus, the functionally important reactions are not necessarily lethal, and this point places a limitation to the reaction-centric approach with studying lethality by observing the gene deletion effects. In this regard, we have investigated the interplay between cellular robustness and the underlying metabolism from the metabolite point of view, and how the robustness can be accomplished at the level of the metabolites, which are the fundamental entities (4, 8) generated, consumed, and recycled by the metabolic processes. Constraints-based flux analysis was carried out under various genotypic and environmental conditions by using the genome-scale *Escherichia coli* metabolic model consisting of 762 metabolites and 932 biochemical reactions (9–13) (see *Materials and Methods*).

**Characterization and *in vivo* Validation of Metabolite Essentiality.** To explore the robustness of *E. coli* metabolism from the metabolite perspective, we first classified all intracellular metabolites into two categories, essential and non-essential metabolites, by monitoring cell growth when the consumption rate of a given metabolite is suppressed to zero (*Materials and Methods*). The resultant list of essential metabolites is given in supporting information (SI) Table 1 under 19 different environmental conditions specified by different combinations of several C, P, N, and S sources, and aerobic vs. anaerobic conditions (SI Table 2). The results obtained in glucose-minimal medium under aerobic condition were used as the representative examples. Interestingly, the relatively unstudied metabolites, hexadecenoyl-ACP, phosphatidylglycerol, and 2-isopropylmaleate, were found to be essential. The metabolite essentiality does not depend much on the environmental conditions because 87.8% of total essential metabolites are always essential for different growth conditions (Fig. 1*a*).

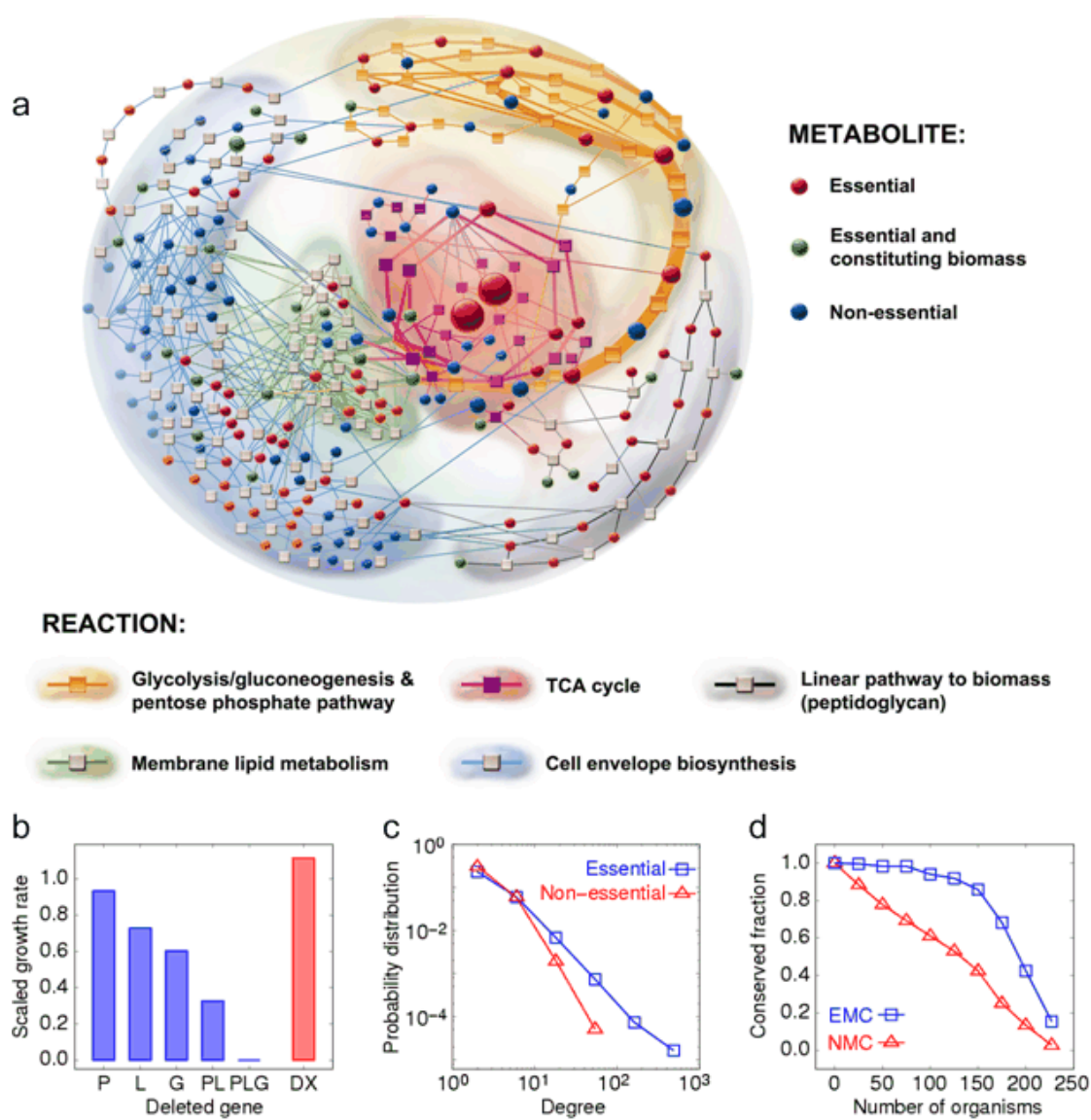

**Fig. 1.** Characteristics of essential and non-essential metabolites in *E. coli* metabolism. (*a*) Metabolic network including the central and the cell envelope metabolism. Cofactors are not shown here because the number of the associated reactions is too large for visual examination. The size of each circle/box corresponds to the amount of flux associated with a metabolite/reaction, whereas thickness of each line denotes the flux

across the line. (*b*) Experimental measure of growth rate relative to that of wild type after disrupting the genes around an essential metabolite tetrahydrofolate (blue), or around a non-essential metabolite 1-deoxy-D-xylulose 5-phosphate (red). The deleted genes: P, *purN*; L, *lpdA*; G, *glyA*; PL, *purN*/*lpdA*; PLG, *purN*/*lpdA*/*glyA*; DX, *dxs*/*xylB*. (*c*) Distributions of degree *k* for essential and non-essential metabolites. The vertical axis represents *P(k)* defined as fraction per degree, $\int_{k-\delta}^{k+\delta} P(k')dk' = f(k)$ where *f(k)* is the fraction of metabolites between *k*−*δ* and *k*+*δ*. Such distributions follow a power-law $P(k) \propto k^{-\gamma}$ with $\gamma = 1.97$ for essential metabolites and with $\gamma = 3.06$ for non-essential ones. (*d*) The horizontal axis represents the number of different organisms *N*, whereas the vertical axis represents the fraction of metabolites conserved phylogenetically in >*N* different organisms. EMC/NMC denotes the metabolites of *E. coli* essential/non-essential for more than half of growth conditions. The majority of EMC (66.1%) are present in most of the organisms (≥79.3%), contrary to the case of NMC (only 21.2% in the same phylogenetic range).

---

The essentiality of a given metabolite can be demonstrated *in vivo* by means of multiple gene disruptions around the metabolite. If disrupting the multiple non-lethal reactions around a particular metabolite suppresses cell growth, the metabolite can be regarded as essential because the deletion of the individual reaction itself is already non-lethal. We conducted the gene deletion experiments for the neighboring eight reactions around tetrahydrofolate, which was identified as an essential metabolite *in silico*. Among these reactions (genes), phosphoribosylglycinamide formyltransferase (*purN*), glycine cleavage system (*lpdA*), glycine hydroxymethyltransferase (*glyA*) were selected as disruption targets. Each single and double gene deletion mutant (Δ*purN*, Δ*lpdA*, Δ*glyA*, and Δ*purN*Δ*lpdA*) was able to survive although with some growth rate changes, but the simultaneous deletion of all of the three genes (Δ*purN*Δ*lpdA*Δ*glyA*) prevented cell growth completely, indicating that the tetrahydrofolate is indeed essential for cell growth (Fig. 1*b*). In contrast, 1-deoxy-D-xylulose 5-phosphate was identified as a non-essential metabolite *in silico*, and experimental disruption of all of the reactions producing the metabolite by constructing Δ*dxs*Δ*xylB* only slightly changed or even increased the growth rate compared with the wild type (Fig. 1*b*). These results indicate that multiple gene knockout mutants for the reactions around essential metabolites can suffer from the detrimental impact on cellular function, whereas those around non-essential metabolites have a negligible influence on growth capability. Throughout these experiments, the measured growth rates of the gene deletion mutants relative to that of the wild type were consistent with the *in silico* predictions (Note 1 in supporting information).

**Structural and Evolutionary Properties of Metabolites.** We also investigated the inherent network properties of essential metabolites to elucidate the correlation between the structural property and functional behavior from the metabolite perspective. First, the number of reactions (degree) participated in by each metabolite was calculated (5). The degree distributions for both essential and non-essential metabolites were found to follow a power-law distribution over the broad range of degrees (Fig. 1*c*). Not surprisingly, the degree distribution of essential metabolites is more right-skewed compared with non-essential ones, indicating that essential metabolites are connected with more reactions than non-essential ones. Indeed, most of highly connected metabolites are essential; they mostly include essential molecules and cofactors, i.e., H, $H_2O$, ATP, Pi, ADP and NAD, which participate in >76 reactions. Among the metabolites having the degree of <3, only 34% of them were found to be essential. It should be noticed that many of non-essential metabolites manifest inactive state where all fluxes from and to such metabolites remain zeros. If these inactive metabolites among non-essential metabolites are not considered, 88.6% of the active metabolites having the degree of <3 become essential.

Because the loss of essential metabolites directly threatens cell viability, one would expect that the metabolites that are essential under various growth conditions should be well conserved across species. We investigated the evolutionary conservation of the essential metabolites in 227 organisms with fully sequenced genomes (SI Table 3). Indeed, the metabolites essential for most growth conditions of *E. coli* were present in many different organisms, showing a much higher degree of conservation than the non-essential ones during the evolutionary process (Fig. 1*d*).

**Stability of Metabolite Flux-Sum.** To understand the robustness of the cellular metabolism quantitatively from the metabolite perspective, the strength of all fluxes in and out of each metabolite was quantified. To this end, the flux-sum (Φ) of the metabolite was defined as the summation of all incoming or outgoing fluxes as follows:

$$\Phi_i = \sum_{j \in P_i} S_{ij} v_j = -\sum_{j \in C_i} S_{ij} v_j = \frac{1}{2} \sum_j \left| S_{ij} v_j \right|$$

where $S_{ij}$ is the stoichiometric coefficient of metabolite $i$ in reaction $j$, and $v_j$ is the flux of reaction $j$. $P_i$ denotes the set of reactions producing metabolite $i$, and $C_i$ denotes the set of reactions consuming metabolite *i*. Under the pseudo-steady state assumption, $\Phi_i$ is the mass flow contributed by all of the fluxes producing (consuming) metabolite *i*.

The robustness of *E. coli* metabolism was examined by determining the sensitivity of the flux-sum to genetic perturbation around a given metabolite. It was quantified by evaluating the relative fluctuation of $\Phi_i$ in response to deletion of active non-lethal reactions: $\sqrt{\langle \Phi_i^2 \rangle - \langle \Phi_i \rangle^2}/\langle \Phi_i \rangle$ where $\langle \cdots \rangle$ denotes the average over each deletion of active non-lethal reactions. It should be noted that we are not interested in those trivial cases with the deletion of inactive reactions. At low relative fluctuation values, the number of essential metabolites was much greater than that of non-essential metabolites (Fig. 2*a*). This result indicates that the flux-sums of essential metabolites are relatively insensitive to genetic perturbation compared with those of non-essential ones. Indeed, 94.3% of total metabolites found in the fluctuation range of <0.0875 are all essential, and there are only non-essential metabolites in the twenty highest ranked ones in relative fluctuations. Thus, it can be concluded that essential metabolites are resistant to internal variations by maintaining the basal mass flow of the corresponding metabolite, thereby leading to the robustness of cellular metabolism.

What mechanism might contribute to such resistance of essential metabolites to internal perturbations? To explicitly tackle this question, we monitored the individual flux values around essential metabolites under genetic perturbations. We defined the flux-vector ($\vec{\Psi}$) of metabolite *i* as a collection of individual fluxes $S_{ij}v_j$ for all its linked reactions *j*, $\vec{\Psi}_i = \{S_{ij}v_j\}$, and evaluated the flux-vector fluctuation, which represents the relative deviation of the flux values around the given metabolite upon deleting reactions: $\sqrt{\langle |\vec{\Psi}_i|^2 \rangle - |\langle \vec{\Psi}_i \rangle|^2}/|\langle \vec{\Psi}_i \rangle|$ where $|\cdots|$ denotes the magnitude of a given vector. Apparently, the observed variation of relevant fluxes around the metabolite directly contributes to the change of the flux-sum for the metabolite. Scatter plot between the flux-sum fluctuation and the corresponding flux-vector fluctuation for non-essential metabolites clearly shows a linear relationship, indicating that the flux-sum of the metabolites is mostly affected by perturbed variations in individual fluxes (Fig. 2*b*). On the other hand, a considerable number of essential metabolites show only small fluctuation in their flux-sums despite of the increased fluctuations in the individual flux values (Fig. 2*b*). This result implies that the flux-sums of essential metabolites are not much affected by the flux variations around them, compared with those of non-essential ones.

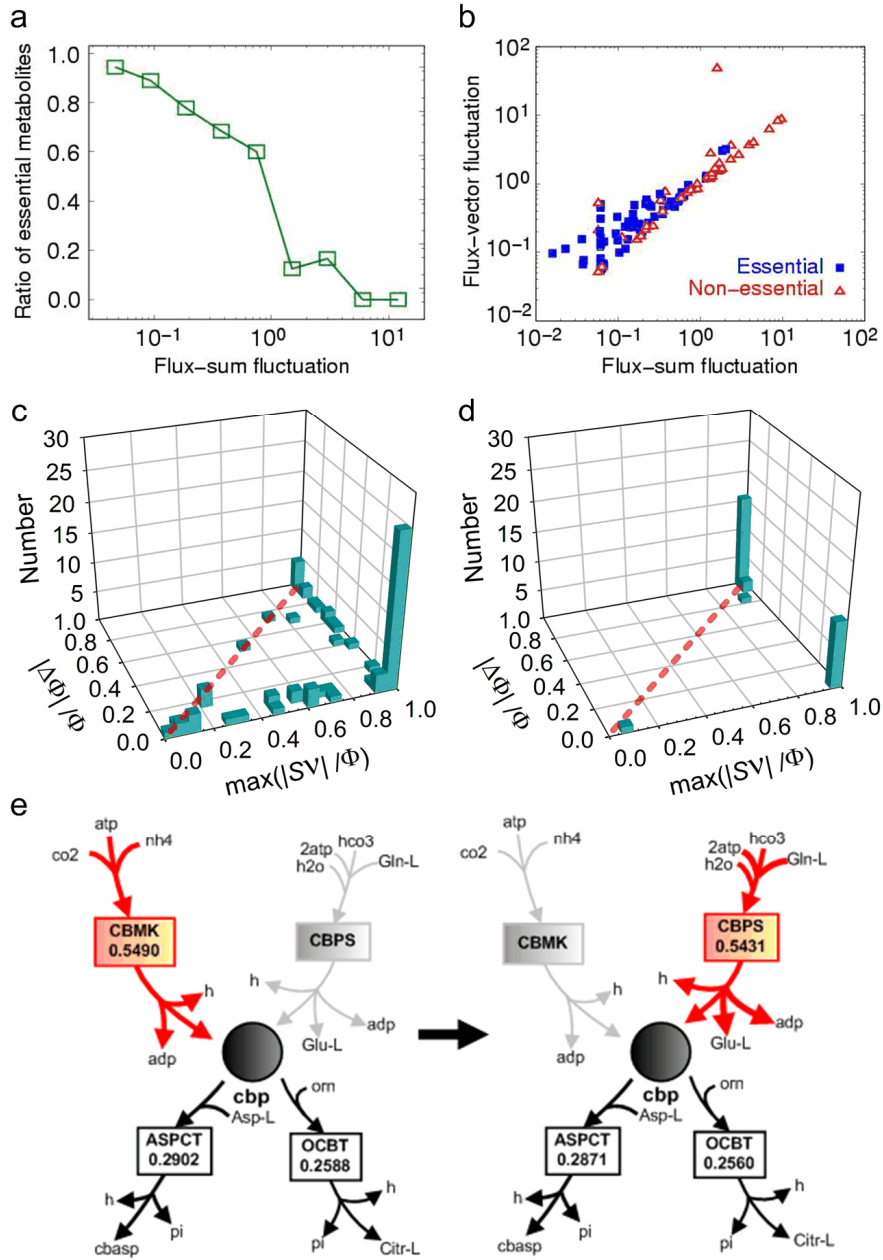

**Fig. 2.** Stability of metabolite flow under genetic perturbations. (*a*) The ratio of essential metabolites to all metabolites as a function of flux-sum fluctuation. (*b*) Flux-sum fluctuation versus flux-vector fluctuation for each of essential/non-essential metabolites. (*c* and *d*) The number of essential metabolites (*c*) or non-essential ones (*d*) in which metabolite $i$ takes the relative change of flux-sum $|\Delta \Phi_i| / \Phi_i$ when the active non-lethal reaction $j$ with the maximum contribution to the flux-sum (maximum of $|S_{ij}v_j|$

/$\Phi_i$) is removed. The cases below (over) the diagonal indicate the additional flux compensation (loss) from other reactions. (*e*) Illustration of the neighboring reactions for carbamoyl phosphate (cbp). The flux-sum of carbamoyl phosphate becomes compensated for by carbamoyl-phosphate synthase (CBPS) when the reaction of the highest flux, carbamate kinase (CBMK), is removed. Thickness of each arrow represents the amount of flux, as shown below the name of the reaction.

---

To clarify such resistance of essential metabolites against internal perturbations, severe perturbation was conducted by deleting the reaction which contributes most to the flux-sum of a given essential metabolite. It should be noted that the deleted reaction is an active non-lethal reaction linked to the metabolite. Fig. 2 *c* and *d* shows the effects of this kind of severe perturbation on the flux-sum changes of essential and non-essential metabolites, respectively. Most essential metabolites are located below the diagonal, indicating that the extent of flux-sum change is less than the flux loss caused by deleting the most contributing reaction. Accordingly, even though the reaction having a relatively high flux is eliminated, the flux-sum can be compensated for by other fluxes around the essential metabolite, recovering such flux loss immediately. Remarkably, for many essential metabolites, the flux loss can be mostly recovered by the fluxes of other reactions, thereby resulting in a very small change of the flux-sum, even when the dominant reaction with a very high flux value (6) is removed (Fig. 2*c*). Such metabolites include carbamoyl phosphate, dUMP, CMP, and L-glutamate 5-semialdehyde (Note 3 in supporting information). For example, carbamoyl phosphate is a key metabolite involved in arginine and proline metabolism and in purine and pyrimidine biosynthesis. The flux-sum of carbamoyl phosphate could be maintained by alteration of other fluxes when the largest flux of the reaction catalyzed by carbamate kinase is blocked completely (Fig. 2*e*); it was found that carbamoyl-phosphate synthase could compensate for the large flux loss caused by knocking-out carbamate kinase, resulting in the recovery of 98.9% of the basal flux-sum. The up-regulation of carbamoyl-phosphate synthase in response to the deletion of carbamate kinase is actually inferred from the gene expression profile data (Note 3 in supporting information). These results suggest that the maintenance of the flux-sum can serve as a good indicator of metabolic robustness. This fact motivated us to predict efficiently the candidate reactions being activated for the flux-sum recovery under the severe gene knockout perturbations. Indeed, using the stoichio-similarity, we developed an algorithm to predict the most probable reaction which would recover the flux-sum after the gene knockout perturbation (Note 3 in supporting information). Therefore, we believe that cellular robustness can be elucidated by such functional property of the metabolic network manifesting the resilience of essential metabolites against the disturbed flux conditions.

**Attenuation of Metabolite Flux-Sum.** Essential metabolites play a pivotal role in cell survival, steadily maintaining the mass flow to produce or consume the metabolites against any internal disturbances within the cell. In other sense, this metabolite perspective on the robustness of *E. coli* provides cellular-level fragility: the failure to maintain the flux-sum of a single essential metabolite can drastically suppress whole cellular growth. The malfunction of multiple genes around the metabolite might cause such critical decrease in the flux-sum. Especially, for most essential metabolites (85%), reducing the flux-sum by half of the basal level led to a suppression of the growth rate by one half or more, whereas only 28.9% of active non-essential metabolites showed such behavior.

The effects of reducing the flux-sum on cell growth were examined next. When the flux-sum was gradually decreased, each essential metabolite exhibited a characteristic profile of the cell growth rate, which belonged to one of three types – A, B, and C, as in Fig. 3$a^*$. The growth rate was sensitive to the extent of flux-sum reduction for types A and C, but not so much for type B. Such characteristic of the growth rate seems to be correlated with the basal flux-sum values; the metabolites of type A had low basal flux-sums, those of type B had high basal flux-sums, and those of type C had ultra-high basal flux-sums (Fig. 3 *b* and *c*). It turns out that 83.8% of essential metabolites belong to type A. These metabolites adjust the cell growth rate proportionally to the flux-sum, and thus act like acclimators affecting the cell growth; the acclimator metabolites allow the cell growth rate to be finely adjusted through their flux-sums, and thereby provide an effective control of cell growth. The classification of essential metabolites according to the growth profile under flux-sum attenuation is described in detail in Note 4 of supporting information.

---

\* Exceptionally, ubiquinol-8, ubiquinone-8, and L-malate can exhibit the different growth profile termed as type D. For more details, refer to Note 4 in supporting information.

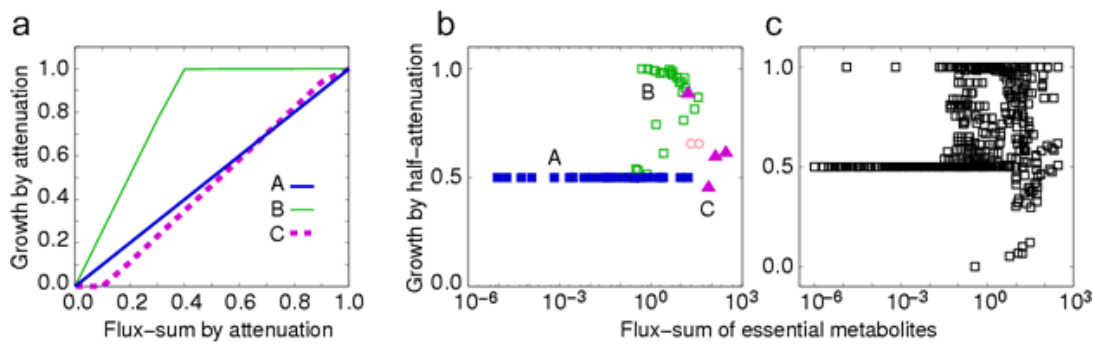

**Fig. 3.** Phenotypic effect by attenuating the flux-sum level of essential metabolites. (*a*) Changes of cell growth rate as the flux-sum continuously decreases. The growth rate and flux-sum are scaled relative to those of wild type. The growth profile of type A is for the case of the metabolite, phosphatidylglycerophosphate, that of type B for oxidized thioredoxin, and that of type C for adenosine diphosphate. (*b* and *c*) Cell growth rate relative to the wild type by reducing the flux-sum to a half, shown as a function of basal flux-sum level of attenuated essential metabolites. The results from the glucose-minimal aerobic condition (*b*) and from all 19 environmental conditions (*c*) are presented. The metabolites in *b* are colored in the same way as in *a* according to the growth profile. For the metabolites that do not belong to any type in *a*, refer to Note 4 in supporting information.

## Discussion

The functional robustness of metabolic networks is the outcome of a long evolutionary process that reflects the resistance toward internal and external fluctuations (14–17). For example, the existence of alternative pathways or flux redistributions implies that these backup pathways might possibly be activated to perform the same function under various genetically and environmentally perturbed conditions (3, 18). Such fault-tolerance or robustness might be a key to cell survival against these perturbations. In this regard, a metabolite-based perspective can provide new guidelines for interpreting cellular robustness. Essential metabolites substantial to cell survival are capable of rerouting metabolic fluxes while sustaining their usage level. This capability of the essential metabolites leads to quite dramatic tolerance to a wide range of internal disturbances. It is possible that some essential metabolites may not show the characteristics presented here. This inaccuracy may arise because regulatory mechanisms have not been fully considered in the current analysis. We have only examined the effects of incorporating a limited number of regulatory mechanisms during this study, which have not shown much difference (Note 1 in supporting information). It is expected that better classification of essential and non-essential metabolites can be performed when genome-wide regulatory mechanisms are incorporated in the genome-scale flux analysis.

A number of applications can be envisaged by using the concept of metabolite essentiality. For example, it can be used to develop metabolic engineering strategies for enhanced production of desired bioproducts by suitably implementing the desired flux values. It can also be used to identify new drug targets. Disruption (knockout) of multiple non-lethal reactions (genes) around an essential metabolite can lead to fatal cell damage and even the attenuation (knockdown) of those reactions might have the same effect. In the case of treating superbacteria that are resistant to multiple antibiotics, one can design drugs that inhibit those enzymes catalyzing multiple non-lethal reactions around an essential metabolite. Alternatively, synthetic lethal mutations (7, 19, 20) can be systematically identified for those enzymes by various screening techniques (21), and implemented by siRNA and other anti-sense techniques.

## Materials and Methods

**Constraints-Based Flux Analysis.** The genome-scale *in silico E. coli* metabolic model iJR904 was used with slight modifications based on the publicly available information and databases (9–12); it consists of 762 metabolites (including external metabolites) and 932 biochemical reactions (including transport processes). Cell growth was quantified by a biomass equation derived from the drain of biosynthetic precursors and relevant cofactors into *E. coli* biomass at their appropriate ratios (13). The stoichiometric relationships among all metabolites and reactions of the genome-scale *in silico E. coli* model were balanced under the steady-state hypothesis. The resultant balanced reaction model is, however, almost always underdetermined in calculating the flux distribution because of insufficient measurements and/or constraints. Thus, the unknown fluxes within the metabolic reaction network were calculated by linear programming-based optimization with an objective function of maximizing the growth rate, subject to the constraints pertaining to mass conservation, reaction thermodynamics, and capacities as follows: $\sum_{j \in J} S_{ij} v_j = b_i$, $\alpha_j \leq v_j \leq \beta_j$, where $S_{ij}$ represents the stoichiometric coefficient of metabolite $i$ in reaction $j$, $v_j$ the flux of reaction $j$, $J$ the set of all reactions, and $b_i$ the net transport flux of metabolite $i$. If this metabolite is an intermediate, $b_i$ would be zero. $\alpha_j$ and $\beta_j$ are the lower and upper bounds of the flux of reaction $j$, respectively. Herein, the flux of any irreversible reaction is considered to be positive: the negative flux signifies the reverse direction of the reaction. The intracellular fluxes were quantified to elucidate the robustness of *E. coli* metabolism in response to genetic perturbations under various environmental conditions (SI Table 2). We also performed the simulation with additional regulatory constraints (22) and another optimization scheme, MOMA (minimization of metabolic adjustment) (23), and found no qualitative difference from the results presented here.

**Characterization of Metabolite Essentiality.** The metabolite essentiality can be defined as the phenotypic effect of a metabolite *M* on cell growth when its consumption rate is set to zero. All fluxes around the metabolite *M* should be restricted to only produce the metabolite, for which balancing constraint of mass conservation is relaxed to allow nonzero values of the incoming fluxes whereas all outgoing fluxes are limited to zero. As such, other metabolites linked to the reactions producing the metabolite *M* can be consistently taken into account, preventing the phenotypic effect irrelevant to the essentiality of the given metabolite *M*. We scaled the resultant change of cell growth rate relative to the growth rate of the wild type for calculating the essentiality of the metabolite. When all reactions around the metabolite were inactive for specific growth condition, we considered that metabolite as non-essential. Because the essentiality of all metabolites follows a clear bimodal distribution (SI Fig. 5), an essential metabolite can be easily identified when its absence leads to decrease in cell growth rate at least a half of that of the wild type, whereas the absence of a non-essential metabolite has minimal or no effect on cell growth. We also tried other criteria for essential/non-essential metabolites according to this essentiality but did not find much difference.

**Construction of Gene Knockout Mutant Strains.** Mutant strains were constructed by the one-step gene inactivation method (24). The wild-type *E. coli* W3110 strain was transformed with pKD46 that contains phage λ recombination system. *E. coli* W3110 cells carrying pKD46 were transformed by electroporation with a PCR product that was produced by using either plasmid pKD3 or pKD4 as templates. The PCR product had 50- to 56-bp homology to the upstream and downstream DNA immediately adjacent to the specific target gene to be knocked-out, and also contained Flp recombinase target site (FRT). Recombinant strains were selected by growing cells in the presence of chloramphenicol or kanamycin, and the inserted cassette was eliminated by using a helper plasmid pCP20. Each knockout mutant was confirmed by PCR analysis using the primers that were not in the region of the gene deletion. The knockout mutants were grown in Luria-Bertani (LB) broth or on LB agar plates at 37℃.

**Measurement of Specific Growth Rate.** All strains were grown in M9 minimal medium containing 5 g/liter glucose to determine their growth kinetics by using Microbiology Reader Bioscreen C analyzer (Oy Growth Curves AB Ltd, Helsinki, Finland). Detailed procedures are described in Note 1 of supporting information

## Acknowledgements

We thank C.-M. Ghim, S. B. Sohn, H. U. Kim, J. S. Yang, and H. Yun for providing valuable information and linguistic advice and S. T. Kim and J.-H. Pak for illustrative assistance. The pKD46 plasmid used for our gene-disruption experiments was kindly provided by B. L. Wanner. This work was supported by Korean Systems Biology Program Grant M10309020000-03B5002-00000 of the Ministry of Science and Technology and by the LG Chem Chair Professorship. H.J. acknowledges the financial support from KOSEF through Grant R17-2007-073-01001-0.